\documentclass[reprint,twocolumn,final,aps,prb,showpacs,superscriptaddress,amsmath,amssymb,amsfonts,floatfix]{revtex4-2}

\usepackage{graphicx}
\usepackage{bm}
\usepackage{color}
\usepackage{mathrsfs}
\usepackage{textcomp, gensymb}
\usepackage{upgreek}
\usepackage[colorlinks=true,linkcolor=blue,citecolor=blue]{hyperref}

\usepackage{stackrel}
\usepackage{floatrow}

\usepackage{multirow}
\usepackage{url}

\usepackage{tabularx}
\usepackage{sidecap}
\usepackage{soul}
\usepackage{float}
\usepackage{color}
\usepackage{orcidlink}
\begin{document}

\title{Etching-free dual lift-off for direct patterning of epitaxial oxide thin films}

\author{Jiayi Qin}
\affiliation{Faculty of Materials Science and Engineering, Kunming University of Science and Technology, Kunming, 650093, Yunnan, China}

\author{Josephine Si Yu See}
\affiliation{Division of Physics and Applied Physics, School of Physical and Mathematical Sciences, Nanyang Technological University, Singapore 637371, Singapore}

\author{Yanran Liu}
\affiliation{Division of Physics and Applied Physics, School of Physical and Mathematical Sciences, Nanyang Technological University, Singapore 637371, Singapore}

\author{Xueyan Wang}
\affiliation{Division of Physics and Applied Physics, School of Physical and Mathematical Sciences, Nanyang Technological University, Singapore 637371, Singapore}

\author{Wenhai Zhao}
\affiliation{Faculty of Materials Science and Engineering, Kunming University of Science and Technology, Kunming, 650093, Yunnan, China}

\author{Yang He}
\affiliation{Yunnan Key Laboratory of Electromagnetic Materials and Devices, National Center for International Research on Photoelectric and Energy Materials, School of Materials and Energy, Yunnan University, Kunming, 650091, Yunnan, China}

\author{Jianbo Ding}
\affiliation{Faculty of Materials Science and Engineering, Kunming University of Science and Technology, Kunming, 650093, Yunnan, China}

\author{Yilin Wu}
\affiliation{Faculty of Materials Science and Engineering, Kunming University of Science and Technology, Kunming, 650093, Yunnan, China}

\author{Shanhu Wang}
\affiliation{Faculty of Materials Science and Engineering, Kunming University of Science and Technology, Kunming, 650093, Yunnan, China}

\author{Huiping Han}
\affiliation{Faculty of Materials Science and Engineering, Kunming University of Science and Technology, Kunming, 650093, Yunnan, China}

\author{Afzal Khan}
\affiliation{Faculty of Materials Science and Engineering, Kunming University of Science and Technology, Kunming, 650093, Yunnan, China}

\author{Shuya Liu}
\affiliation{Faculty of Materials Science and Engineering, Kunming University of Science and Technology, Kunming, 650093, Yunnan, China}

\author{Sheng'an Yang}
\affiliation{Faculty of Materials Science and Engineering, Kunming University of Science and Technology, Kunming, 650093, Yunnan, China}

\author{Hui Zhang}
\affiliation{Faculty of Materials Science and Engineering, Kunming University of Science and Technology, Kunming, 650093, Yunnan, China}

\author{Jiangnan Li}
\affiliation{Faculty of Materials Science and Engineering, Kunming University of Science and Technology, Kunming, 650093, Yunnan, China}

\author{Qingming Chen}
\affiliation{Faculty of Materials Science and Engineering, Kunming University of Science and Technology, Kunming, 650093, Yunnan, China}
\affiliation{Southwest United Graduate School, Kunming, Yunnan 650500, China}

\author{Jiyang Xie}
\affiliation{Yunnan Key Laboratory of Electromagnetic Materials and Devices, National Center for International Research on Photoelectric and Energy Materials, School of Materials and Energy, Yunnan University, Kunming, 650091, Yunnan, China}
\affiliation{Electron Microscopy Center, Yunnan University, Kunming, 650091, Yunnan, China}

\author{Ji Ma\orcidlink{0000-0002-1308-3226}}
\affiliation{Faculty of Materials Science and Engineering, Kunming University of Science and Technology, Kunming, 650093, Yunnan, China}
\affiliation{Southwest United Graduate School, Kunming, Yunnan 650500, China}

\author{Wanbiao Hu}
\affiliation{Yunnan Key Laboratory of Electromagnetic Materials and Devices, National Center for International Research on Photoelectric and Energy Materials, School of Materials and Energy, Yunnan University, Kunming, 650091, Yunnan, China}
\affiliation{Electron Microscopy Center, Yunnan University, Kunming, 650091, Yunnan, China}

\author{Jianhong Yi}
\affiliation{Faculty of Materials Science and Engineering, Kunming University of Science and Technology, Kunming, 650093, Yunnan, China}
\affiliation{Southwest United Graduate School, Kunming, Yunnan 650500, China}
	
\author{Liang Wu\orcidlink{0000-0003-1030-6997}}
\email{liangwu@kust.edu.cn}
\affiliation{Faculty of Materials Science and Engineering, Kunming University of Science and Technology, Kunming, 650093, Yunnan, China}

\author{X. Renshaw, Wang\orcidlink{0000-0002-5503-9899}}
\email{renshaw@ntu.edu.sg}
\affiliation{Division of Physics and Applied Physics, School of Physical and Mathematical Sciences, Nanyang Technological University, Singapore 637371, Singapore}
\affiliation{School of Electrical and Electronic Engineering, Nanyang Technological University, Singapore 639798, Singapore}

\date{\today}

\begin{abstract}

Although monocrystalline oxide films offer broad functional capabilities, their practical use is hampered by challenges in patterning. 
Traditional patterning relies on etching, which can be costly and prone to issues like film or substrate damage, under-etching, over-etching, and lateral etching.
In this study, we introduce a dual lift-off method for direct patterning of oxide films, circumventing the etching process and associated issues. Our method involves an initial lift-off of amorphous Sr$_3$Al$_2$O$_6$ or Sr$_4$Al$_2$O$_7$ ($a$SAO) through stripping the photoresist, followed by a subsequent lift-off of the functional oxide thin films by dissolving the $a$SAO layer. 
$a$SAO functions as a ``high-temperature photoresist", making it compatible with the high-temperature growth of monocrystalline oxides. Using this method, patterned ferromagnetic La$_{0.67}$Sr$_{0.33}$MnO$_{3}$ and ferroelectric BiFeO$_3$ were fabricated, accurately mirroring the shape of the photoresist.
Our study presents a straightforward, flexible, precise, environmentally friendly, and cost-effective method for patterning high-quality oxide thin films.

\end{abstract}

\maketitle


The application of functional thin films is heavily based on patterning, as exemplified by two of the most widely used technologies: thin-film transistors (TFTs) \cite{Geng2023} and complementary metal-oxide-semiconductor (CMOS) \cite{Bo2007} devices. In TFTs, patterning defines critical structures such as electrodes and semiconductor channels, enabling high-performance displays (e.g., LCDs, OLEDs). For CMOS, advanced patterning techniques (e.g., lithography) are indispensable for the fabrication of nanoscale transistors and interconnects in integrated circuits. Without precise patterning, neither technology could achieve the miniaturization, functionality, or performance demanded by modern electronics. 

Over the last few decades, research interest has grown mainly in functional oxide thin films due to their novel functionalities that either surpass or complement traditional TFT and CMOS technologies \cite{Schlom2008, Rondinelli2011, Ogale2013, Tarancon2019}. However, the complexity involved in patterning epitaxial oxide thin films greatly limits their practical applications. Specifically, oxide thin films are often patterned via either wet or dry etching processes following traditional ultraviolet (UV) lithography (patterning of photoresist by development).

\begin{figure*}[th]
\centering
\includegraphics[width=1\linewidth]{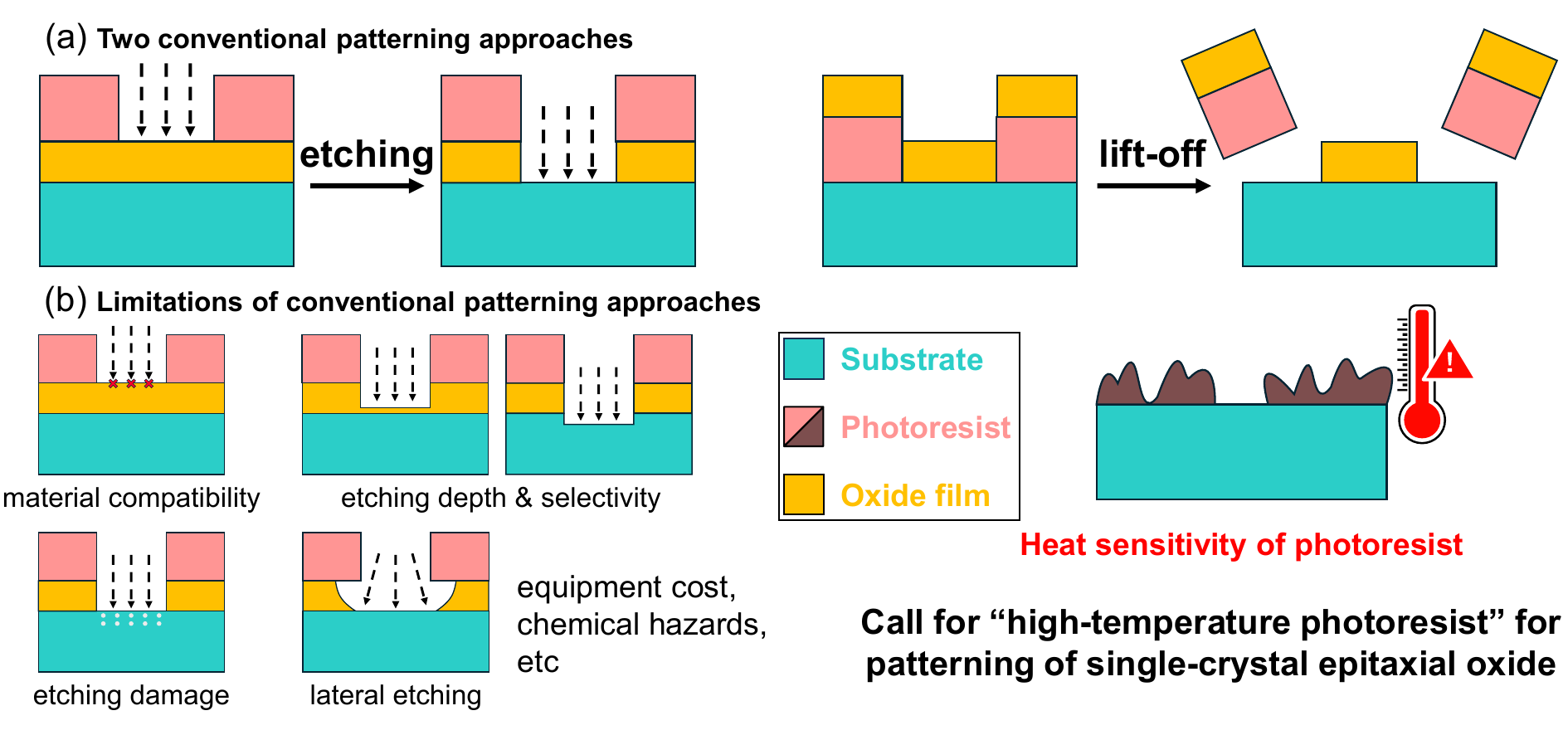}
\caption{(a) Schematic depiction of the two conventional patterning methods: etching (subtractive method) (left) and lift-off (additive method) (right). (b) Limitations of the etching process, which are inherently circumvented by employing the lift-off method. However, exposing the photoresist to high temperatures wuold lead to its deterioration. Consequently, to ensure compatibility with the high-temperature fabrication process, it is essential to develop an alternative to the conventional polymer-based, heat-sensitive photoresist (``high-temperature photoresist").}
\label{Fig1}
\end{figure*}

Etching is a top-down fabrication technique that begins from the surface of the material, removing material to create desired patterns (left panel of schematic \autoref{Fig1}(a)). However, etching processes present several limitations that warrant consideration (left panel of schematic \autoref{Fig1}(b)).
Firstly, material compatibility can be a significant challenge, as certain materials are difficult to etch effectively. For example, Au or Pt are difficult and expensive to remove by etching due to their inertness.
Secondly, achieving precise control over etch depth and material selectivity is challenging, leading to issues such as under-etching, over-etching, and unintended etching of adjacent layers, which can compromise pattern fidelity and layer integrity.
Thirdly, the prevention of lateral etching poses considerable difficulties, which adversely affect the precision and dimensions of the etched features.
Fourthly, it is arduous to prevent etching damage, including ion bombardment damage, resulting in physical damage to the material surface.
Fifthly, many etching techniques utilize hazardous chemicals, necessitating careful handling and disposal practices to ensure safety. The cost and complexity of etching equipment, along with the maintenance required, can also pose significant barriers.
Finally, some etching methods are time-consuming, particularly when high precision is needed \cite{Moreau2012, Madou2011}.

In contrast to subtractive techniques like etching, lift-off is an additive method that employs a sacrificial layer to remove unwanted material from the substrate by detaching the sacrificial layer from the substrate.
Therefore, lift-off avoids these problems, particularly as it minimizes the risk of over-etching, lateral etching, or harming the underlying substrate \cite{Biercuk2003, Banerjee2012, Trung2017, Yu2017}, as shown in the right panel of \autoref{Fig1}(a). For example, the standard commercial method for metal lift-off in patterning metal films involves the following steps: Initially, metallization is typically conducted via physical vapor deposition (PVD), which deposits metal across both the substrate and the photoresist. Organic solvents are subsequently employed as lift-off agents, stripping the remaining denatured photoresist and any attached metal, thus imprinting the resist pattern onto the metal layer \cite{Madou2011}.

However, unlike the room-temperature (RT) deposition of metal layers, monocrystalline epitaxial oxide thin films typically require high-temperature deposition, which can damage the photoresist and render it unsuitable for lift-off processes (right panel of \autoref{Fig1}(b)). Therefore, identifying a photoresist suitable for high-temperature applications or its alternatives is of significant importance.

In this study, we introduce an etching-free dual lift-off technique specifically designed for the direct patterning of oxide films. Initially, analogous to the metal lift-off technique, we achieve the patterning of the amorphous oxides Sr$_3$Al$_2$O$_6$ \cite{Lu2016}  or Sr$_4$Al$_2$O$_7$ \cite{Zhang2024,Nian2024} (SAO) by dissolving the underlying photoresist layer using isopropanol (IPA). In fact, SAO is the primary sacrificial layer for producing freestanding oxide thin films via its non-selective lift-off procedure \cite{Ji2019, Dong2019, Hong2020, Yang2022, Gu2022, Huang2022, Choo2024}, with its amorphous form maintaining high water solubility \cite{Han2019}. 
Here, amorphous Sr$_3$Al$_2$O$_6$ and Sr$_4$Al$_2$O$_7$ are collectively referred to as $a$SAO, with specific designations as $a$SAO326 and $a$SAO427, respectively.
Therefore, the second lift-off step involves patterning the functional oxide thin films by dissolving and eliminating the $a$SAO layer in water.
In other words, the $a$SAO layer acts as the ``high-temperature photoresist".
We illustrate the advanced capabilities and versatility of this patterning technique in preserving film quality by patterning two representative ferroic materials, ferromagnetic La$_{0.67}$Sr$_{0.33}$MnO$_{3}$ (LSMO) \cite{Dagotto2001} and ferroelectric BiFeO$_3$ (BFO) \cite{Liu2022}.
The precision of the oxide pattern in our research is directly correlated with the lithographic resolution of the photoresist. Enhanced exposure accuracy in lithography would correspondingly improve the accuracy of our technique, as the resulting pattern accurately mirrors the shape of the photoresist.
Furthermore, considering that the substrate surface is almost unaltered after detaching the SAO layer, similar to a conventional photoresist \cite{Wang2023, An2024, Shen2025}, this technique is also applicable for fabricating multilayer oxide heterostructures with interlaced stacking configurations.
Our research presents a direct, versatile, high-resolution, eco-friendly, and cost-efficient method for patterning high-quality oxide thin films. 

\begin{figure*}[ht]
\centering
\includegraphics[width=1\linewidth]{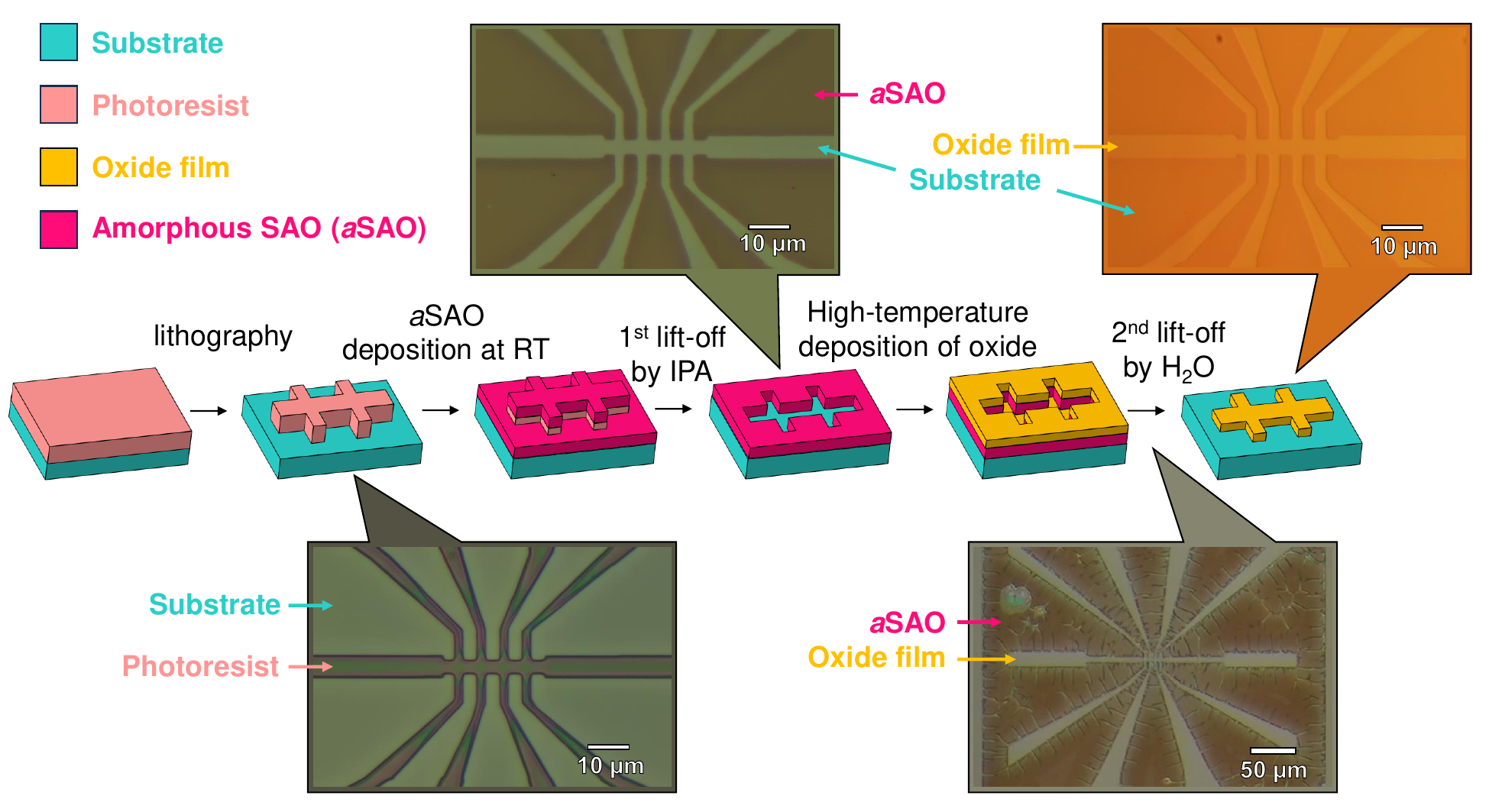}
\caption{Schematic illustration of the etching-free dual lift-off process for patterning epitaxial oxide thin films, along with representative optical microscopy images. The dual lift-off technique involves the following steps: patterning of photoresist, RT growth of $a$SAO, initial lift-off of the photoresist using IPA, high-temperature growth of the oxide film, and final lift-off of SAO using deionized water.}
\label{Fig2}
\end{figure*}

\autoref{Fig2} presents the selective dual lift-off method schematic, accompanied by key optical microscopy images of the pattern structures (the fabrication details are provided in the Supporting Information). As illustrated, the dual lift-off technique was implemented through the following steps: (1) Patterning of the photoresist. (2) Patterning of $a$SAO through an initial lift-off using IPA. (3) Patterning of the thin epitaxial oxide film by a subsequent lift-off with deionized water.

In the first step, a conventional photolithographic process was employed in which a mask with specific structures was exposed to UV light. A positive photoresist (AZ ECI 3012) and a photolithographic developer solution (AZ2026) were used. The thickness of the photoresist measured by atomic force microscopy (AFM) was about 0.82 $\mu$m, as shown in Supporting
Information \autoref{SF1}. It is recommended to keep the photoresist thickness above that of the $a$SAO to avoid forming a continuous film on the photoresist and substrate, which could be crucial for enhancing the edge quality of the remaining $a$SAO after the lift-off of the photoresist.

The second step involves the initial deposition of $a$SAO layer on the SrTiO$_3$ (STO) substrate with the patterned resist, at RT via pulsed laser deposition (PLD) in a background vacuum of approximately $10^{-6}$ Torr. 
As demonstrated in Supporting Information \autoref{SF1}, the thicknesses of $a$SAO326 and $a$SAO427 are kept at 256 nm and 272 nm, respectively, which are less than that of the photoresist.
Following this, the first lift-off (striping) of the photoresist is conducted by immersing the substrate in IPA for 30 minutes, followed by mild ultrasonic cleaning at 40 $\degree$C for 5 minutes to obliterate any remaining photoresist.
It is important to note that acetone is typically employed as the solvent for removing residual photoresist. However, in this case, we used IPA because SAO (amorphous or crystalline) would also dissolve in acetone.
This results in a negative replica of the mask within the $a$SAO layer, leading to a TiO$_2$-terminated STO substrate \cite{Kareev2008} that is coated with $a$SAO featuring patterned openings. 

During the third step, the desired oxide film is epitaxially deposited by PLD (usually at temperatures between 500 and 800 $\degree$C).
In our studies, crystallization of $a$SAO was not observed up to 675 $\degree$C, as shown in Supporting Information \autoref{SF2}; however, even if crystallization were to occur, it would not affect its high solubility in water. Additionally, the high solubility of $a$SAO ensures the efficacy of our dual lift-off method. For instance, an $a$SAO326 layer within a millimeter-scale LSMO/$a$SAO/STO(001) sample can dissolve in merely several tens of minutes at RT, see Supporting Information \autoref{SF2}. To obtain the final patterned oxide film, the patterned $a$SAO sacrificial layer was stripped by immersing it in deionised water for 60 minutes to ensure a complete lift-off and then subjected to ultrasonic agitation at 40 $\degree$C for 5 minutes.

\begin{figure*}[ht]
\centering
\includegraphics[width=\linewidth]{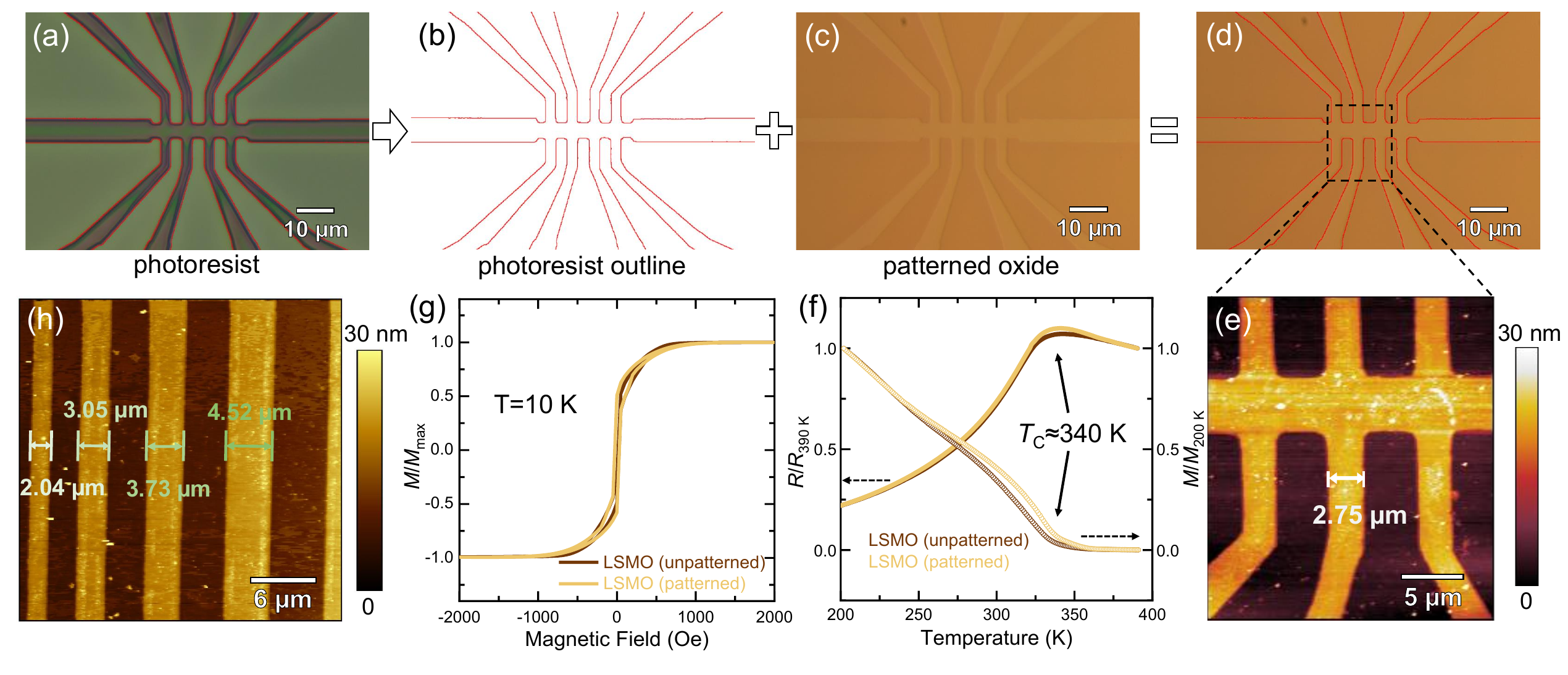}
\caption{(a) Optical microscopy image of the patterned photoresist, the red line represents the pattern outline, which is extracted and shown in (b). (c) Optical microscopy image of the patterned oxide film fabricated using the dual lift-off technique. (d) Overlay of (b) and (c), demonstrating the precise congruence of the outline with the patterned oxide film. (e) AFM image of a Hall-bar structure of LSMO grown on STO, with a scale bar of 5 $\mu$m, it is corresponding to the black-boxed region within the optical microscopy image of the patterned LSMO film shown in (d). Comparison of $R/R_\text{390 K}$--$T$ and $M/M_\text{200 K}$--$T$ curves (f), as well as the $M/M_\text{max}$--$H$ loops (g) for unpatterned and patterned LSMO. A concurrent metal-insulator and ferromagnetic-paramagnetic transition at $T_\text{C} \approx 340$ K, indicating that the dual lift-off process does not deteriorate the properties of the patterned LSMO film.  (h) Patterned LSMO thin films with varying line widths.}
\label{Fig3}
\end{figure*}

Next, we demonstrate the patterning of LSMO thin films using $a$SAO326 as the sacrificial layer.
\autoref{Fig3}(a-d) present a comparison of optical microscopy images between the patterned photoresist and the obtained patterned LSMO film fabricated via our dual lift-off technique. 
The red lines in \autoref{Fig3}(a) represent the outline of the patterned photoresist, as extracted and displayed in \autoref{Fig3}(b).
\autoref{Fig3}(c) presents the optical microscopy image of the resultant oxide pattern, and its combination with \autoref{Fig3}(b) results in \autoref{Fig3}(d).
It is evident that the profile of the photoresist closely matches the outline of the resulting oxide pattern, indicating the high resolution of our dual lift-off technique. The resolution of this patterning method is governed by the precision of the UV photolithography process, meaning the second lift-off procedure does not compromise the resolution of the UV photolithography.

\autoref{Fig3}(e) provides the AFM topography of epitaxial LSMO films ($\sim$ 20 nm, Supporting Information \autoref{SF1}) structured with Hall-bar configurations, illustrating the area highlighted by the black box within the optical microscopy image of the patterned LSMO film shown in \autoref{Fig3}(d). The LSMO pattern possesses very straight edges, maintaining the initially designed line width of 2.75 $\mu$m.
As the oxide thin films were eventually deposited on areas from which the photoresist had been removed, the initial lift-off process facilitates the renewal of the substrate surface, thereby ensuring the patterned films are of equal quality to the continuous unpatterned films.
To confirm this, the temperature-dependent normalized resistivity ($R/R_\text{390 K}$--$T$) and normalized magnetization ($M/M_\text{200 K}$--$T$) of the patterned and unpatterned LSMO thin films deposited on STO substrates under the same conditions were assessed using a physical property measurement system (PPMS, Quantum Design) and magnetic property measurement system (MPMS, Quantum Design).
As illustrated in \autoref{Fig3}(f), these curves obtained from both the patterned and unpatterned LSMO samples exhibit very similar behavior, showing a simultaneous metal-insulator and ferromagnetic-paramagnetic transition at approximately 340 K. This high Curie temperature ($T_\text{C}$) indicates the high-quality LSMO after patterning.
A similar comparison of the $M/M_\text{max}$--$H$ loops further supports the above argument, see \autoref{Fig3}(g).
Thus, the consistent high quality observed in both the patterned thin film and the as-grown unpatterned film suggests that our dual lift-off technique effectively transfers the patterned structures onto thin films while precisely preserving their intrinsic superior quality.
In addition, as demonstrated in \autoref{Fig3}(h), the fabricated LSMO thin films exhibit well-defined patterned structures with varying linewidths, achieving a resolvable feature size of approximately 2 $\mu$m. The constraint on the minimum linewidth in our experiments is due to the maskless UV lithography, rather than the dual lift-off method.

\begin{figure*}[ht]
\centering
\includegraphics[width=.75\linewidth]{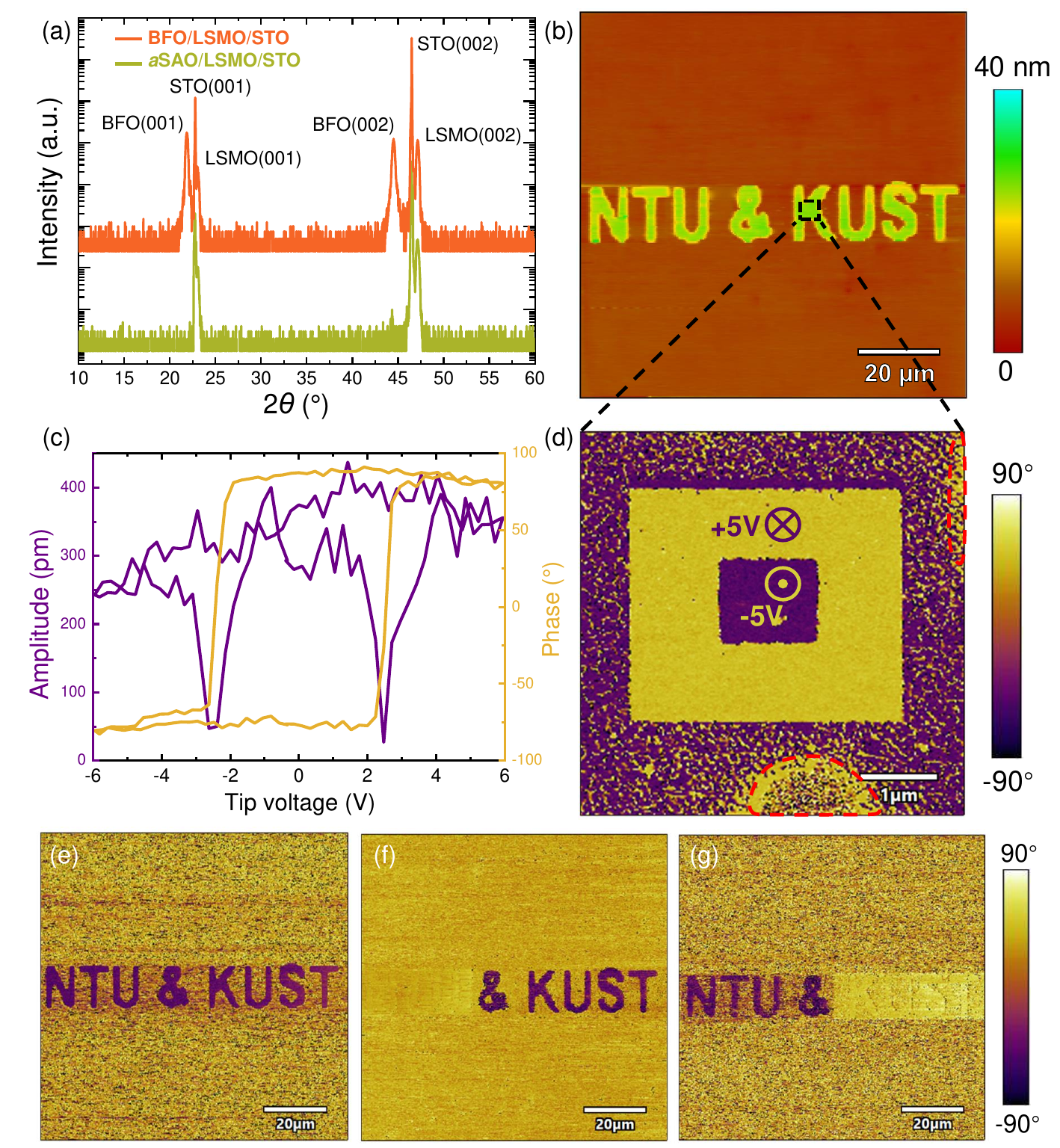}
\caption{(a) XRD 2$\theta$--$\omega$ scans of $a$SAO/LSMO/STO(001) and BFO/LSMO/STO(001) heterostructures. (b) AFM image of the patterned BFO/LSMO/STO thin film. (c) PFM phase and amplitude loops of the patterned BFO thin film. (d) PFM image of the black-boxed area in the AFM image with a $3 \times 3 \ \mu$m$^2$ area poled at $+5$ V ($P_d$) and a nested $1 \times 1 \ \mu$m$^2$ area poled at $-5$ V ($P_u$). The yellow contrast indicates downward polarization, and the purple contrast indicates upward polarization. Note that the ``imperfect'' areas (red dashed area) near the lower edge and upper right corners represent the edge of the letter ``K". (e-g) PFM images depicting as-grown preferred polarization state (e), after ``NTU" and ``KUST" with $+5$ V ($P_d$) and $-5$ V ($P_u$) poling, respectively (f), as well as the images with reversed poling voltage (g).}
\label{Fig4}
\end{figure*}

To showcase the versatility of our patterning method, we additionally produced patterned BFO on LSMO/STO(001). Specifically, a continuous LSMO thin film was first deposited on STO(001) substrate to function as the bottom electrode. Subsequently, a patterned BFO ($\sim$ 19 nm, Supporting Information \autoref{SF1}) in the shape of ``NTU\&KUST" was achieved through $a$SAO427-assisted dual lift-off procedure.
\autoref{Fig4}(a) shows the X-ray diffraction (XRD) 2$\theta$--$\omega$ scans of the $a$SAO427/LSMO/STO(001) and BFO/LSMO/STO(001) heterostructures. It is worth noting that, to achieve sufficiently strong XRD peaks for BFO, the sample prepared for XRD measurements was coated with photoresist in the initial phase of the dual lift-off process, without any exposure. As seen, only STO(00$l$), LSMO(00$l$)$_\text{pc}$, and BFO(00$l$)$_\text{pc}$ (pc denotes pseudocubic) peaks were detected, with no evidence of $a$SAO peaks prior to the deposition of BFO. 
The out-of-plane lattice constants of BFO and LSMO, assessed via Bragg's law, were found to be 3.98 and 3.85 \AA, respectively, aligning with values reported in the literature. \cite{Liu2022}. 

\autoref{Fig4}(b) exhibits AFM images of the fabricated heterostructure, with the ``NTU\&KUST” pattern clearly resolved.  \autoref{Fig4}(c) presents the polarization switching behavior of the patterned BFO film, as characterized by piezoresponse force microscopy (PFM) hysteresis loop measurements.  The results demonstrate 180$\degree$ polarization reversal, exhibiting a square-shaped phase hysteresis loop and a typical butterfly-shaped piezoresponse amplitude loop, with an amplitude of approximately 400 pm and a coercive voltage of approximately 2$\sim$3 V.
To further investigate the ferroelectric properties of the patterned BFO, localized PFM measurements were conducted on the ``K" region. As shown in \autoref{Fig4}(d), a $3 \times 3 \ \mu$m$^2$ area of BFO film was polarized downward ($P_d$ state) by applying $+5$ V DC bias through a conductive probe. Subsequent application of $-5$ V bias achieved nested $1 \times 1 \ \mu$m$^2$ upward polarization ($P_u$ state) within the same region. The distinct 180$\degree$ phase contrast in PFM amplitude images confirms bipolar reversible ferroelectric domain manipulation, indicating the preservation of intrinsic ferroelectric properties despite patterning processes. It should be noted that the as-grown state of BFO exhibits an initial preferred out-of-plane polarization. This is formed during cooling after growth, influenced by boundary conditions such as the physical properties of the specific bottom electrode \cite{Zhang2023}.
We additionally manipulate the ferroelectric polarization across the entire area of ``NTU\&KUST". \autoref{Fig4}(e) displays the PFM phase image of the BFO/LSMO/STO heterostructure without external bias, reflecting the initial polarization state. Subsequently, we applied different DC biases to distinct regions to explore their polarization switching characteristics. Specifically, a $+5$ V bias was applied to the ``NTU" region to switch its polarization direction to downward ($P_d$). Simultaneously, a $-5$ V bias was applied to the ``KUST" region to switch its polarization direction to upward ($P_u$). 
Owing to the aforementioned initial preferred polarization, the ``KUST" retains its upward polarization.
When reversed biases were applied to the ``NTU" and ``KUST" areas, the contrasts also inverted respectively, as shown in \autoref{Fig4}(g). 

In summary, we present a simple, versatile, high-precision, economical, and environmentally friendly approach to the structuring of epitaxial high-quality oxide thin films using an etching-free dual lift-off technique. This involves an initial lift-off of the photoresist to obtain a patterned $a$SAO, followed by a second lift-off of the $a$SAO to achieve a patterned oxide thin film.

\bigskip

\noindent
\textbf{Acknowledgements}

We express our gratitude to Dr. Jiachang Bi, Prof. Yanwei Cao, and Prof. Shuai Ning for their indispensable assistance and productive discussions. L.W. acknowledges funding through the Xingdian Talent Support Project of Yunnan Province (Grant No. KKRD202251009) and Yunnan Fundamental Research Projects (Grant No. 202201AT070171). 
X.R.W. acknowledges support by the Singapore Ministry of Education (MOE) Academic Research Fund (AcRF) Tier 3 grant (MOE-MOET32023-0003) and ``Quantum Geometric Advantage" Tier 1 (Grant No. RG82/23 and RG155/24). 



\bibliographystyle{apsrev4-2}
\bibliography{ref}

\clearpage            
\onecolumngrid

\setcounter{figure}{0}
\setcounter{figure}{0}
\renewcommand{\thefigure}{S\arabic{figure}}
\renewcommand{\theHfigure}{S\arabic{figure}} 

\subsection{Supporting Information}
\onecolumngrid

\noindent
\textbf{S1. Sample fabrication and characterizations}

\medskip
\noindent
\textbf{Photolithography:} Photolithographic patterning was performed using AZ ECI 3012 photoresist. The photoresist was spin-coated onto the substrate at 6000 rpm. The coated sample was then prebaked on a hotplate at 110 $\degree$C for 1 minute and 30 seconds, followed by cooling to room temperature (RT). Exposure was carried out using the TuoTuo lithography machine. A post-exposure bake was conducted at 110 $\degree$C for 1 minute and 30 seconds, after which the sample was cooled again to RT. The development was carried out by immersing the sample in AZ2026 developer solution for approximately 13 seconds, followed by rinsing with deionized water for at least 10 seconds.

\noindent
\textbf{Ceramic targets preparation:}
The La$_{0.67}$Sr$_{0.33}$MnO$_{3}$ (LSMO) target was synthesized via a sol-gel method: La(NO$_3$)$_3$\,$\cdot$\,6H$_2$O, Sr(NO$_3$)$_2$, Mn(NO$_3$)$_2$, citric acid monohydrate (C$_6$H$_8$O$_7$\,$\cdot$\,H$_2$O), and ethylene glycol (C$_2$H$_6$O$_2$) were dissolved in deionized water. The solution was continuously heated with stirring until solvent evaporation yielded a transparent gel. The gel was oven dried, then foamed and ground into a powder. The powder was calcined at 600 $\degree$C to remove volatile components, subsequently re-ground, pressed into pellets, and sintered at 1450 $\degree$C for 22 h. For BiFeO$_3$ (BFO) target preparation, Bi(NO$_3$)$_3$\,$\cdot$\,5H$_2$O, Fe(NO$_3$)$_3$\,$\cdot$\,9H$_2$O, citric acid monohydrate, and ethylene glycol were dissolved in deionized water using the same sol-gel protocol. The resulting gel was calcined at 500 $\degree$C, milled, pressed into pellets, and sintered at 900 $\degree$C for 8 h. Sr$_3$Al$_2$O$_6$ targets were fabricated through a solid-state reaction: SrCO$_3$ and Al$_2$O$_3$ powders were mixed, milled, and subjected to three pre-calcination cycles at 1000 $\degree$C (12 h/cycle). The powder was then pressed and finally sintered at 1400 $\degree$C for 24 h. Sr$_4$Al$_2$O$_7$ targets were similarly prepared by mixing Sr(NO$_3$)$_2$ and Al$_2$O$_3$ followed by sintering at 1450$ \degree$C for 20 h. 

\noindent
\textbf{Thin-film deposition:}
SrTiO$_3$ (STO) substrates were chemically etched with aqua regia (HCl$:$HNO$_3=3:1$), and annealed at 1000 $\degree$C for 1 h to obtain the atomically-flat TiO$_2$-terminated surface. The oxide thin films were deposited on STO(001) substrates via pulsed laser deposition (PLD) using a KrF excimer laser (energy density 1.5 J cm$^{-2}$). For the Sr$_3$Al$_2$O$_6$ or Sr$_4$Al$_2$O$_7$ (SAO) layer, growth was carried out at RT under an oxygen pressure of 10$^{-6}$ Torr with a laser repetition rate of 4 Hz. The LSMO bottom electrode layer was deposited at 675 $\degree$C under 75 mTorr oxygen pressure, while maintaining other parameters consistent with SAO (e.g., laser energy density). The BFO layer was grown at 660 $\degree$C under 100 mTorr oxygen pressure with a repetition rate of 5 Hz, followed by $in$-$situ$ annealing at 150 Torr oxygen pressure with a controlled cooling rate of 15 $\degree$C/min.

\noindent
\textbf{Structural characterizations:}
The crystal structure and surface morphology were characterized by X-ray diffraction (XRD, Rigaku SmartLab) and atomic force microscopy (AFM, Asylum Research, Oxford Instruments).  

\noindent
\textbf{Physical property characterizations: }
Piezoresponse force microscopy (PFM, Asylum Research, Oxford Instruments) was employed to characterize the ferroelectric properties of BFO. Temperature-dependent resistivity and magnetization measurements of LSMO were performed using Physical Property Measurement System (PPMS) and Magnetic Property Measurement System (MPSM) (Quantum Design). 

\newpage
\noindent
\textbf{S2. Thickness measurements}

\begin{figure}[h!]
	\includegraphics[width=\linewidth]{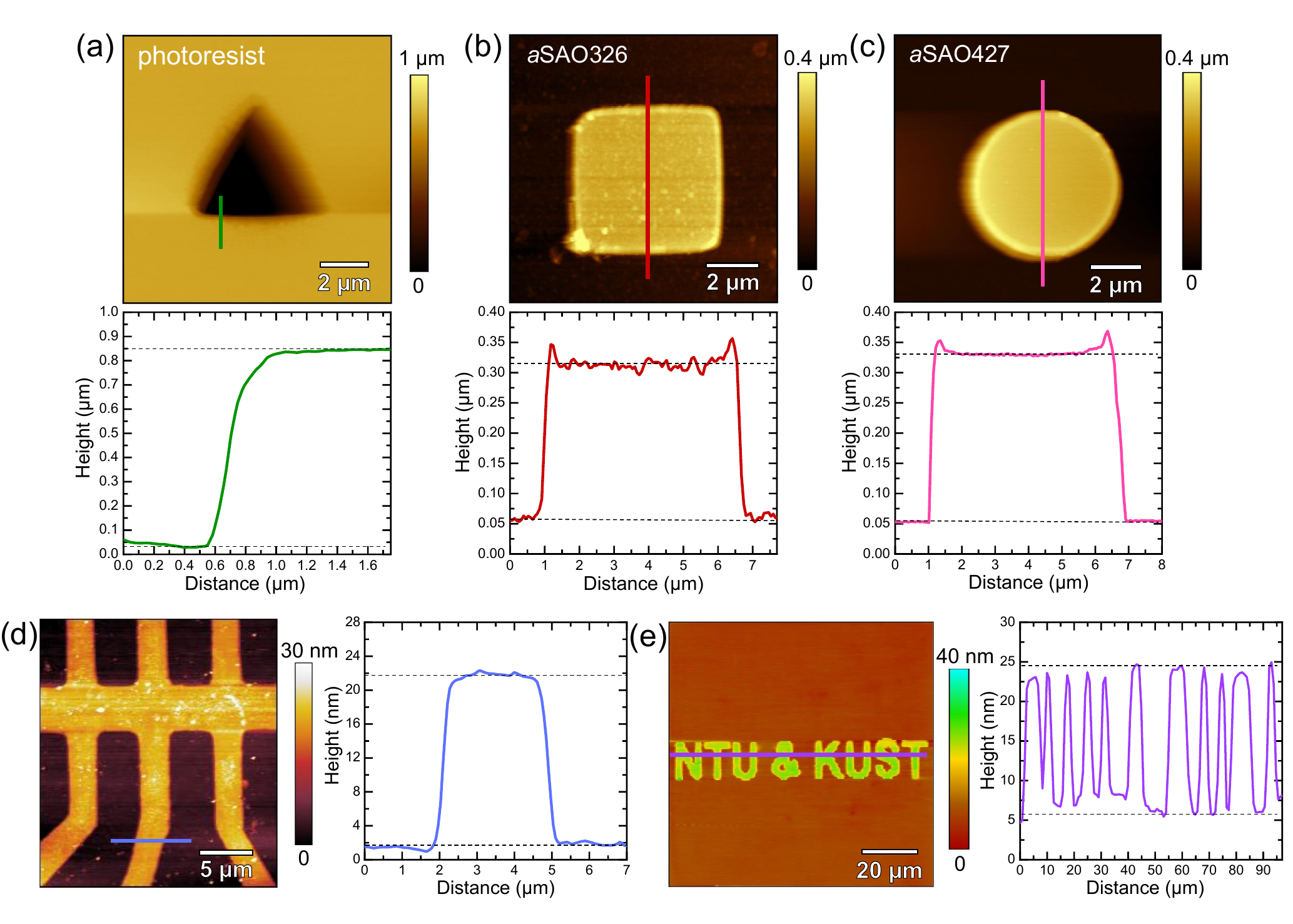}
	\caption{AFM topography images and height profiles for patterned layers of $\sim 0.82 \ \mu$m patterned photoresist (a), $\sim 256$ nm  amorphous Sr$_3$Al$_2$O$_6$ ($a$SAO326) (b), $\sim 272$ nm amorphous Sr$_4$Al$_2$O$_7$ ($a$SAO427) (c), $\sim 20$ nm LSMO (d), and $\sim 19$ nm BFO (e).}
\label{SF1}
\end{figure}

\noindent
\textbf{S3. Structure and solubility of \textbf{\textit{a}}SAO}

\begin{figure}[h!]
	\includegraphics[width=\linewidth]{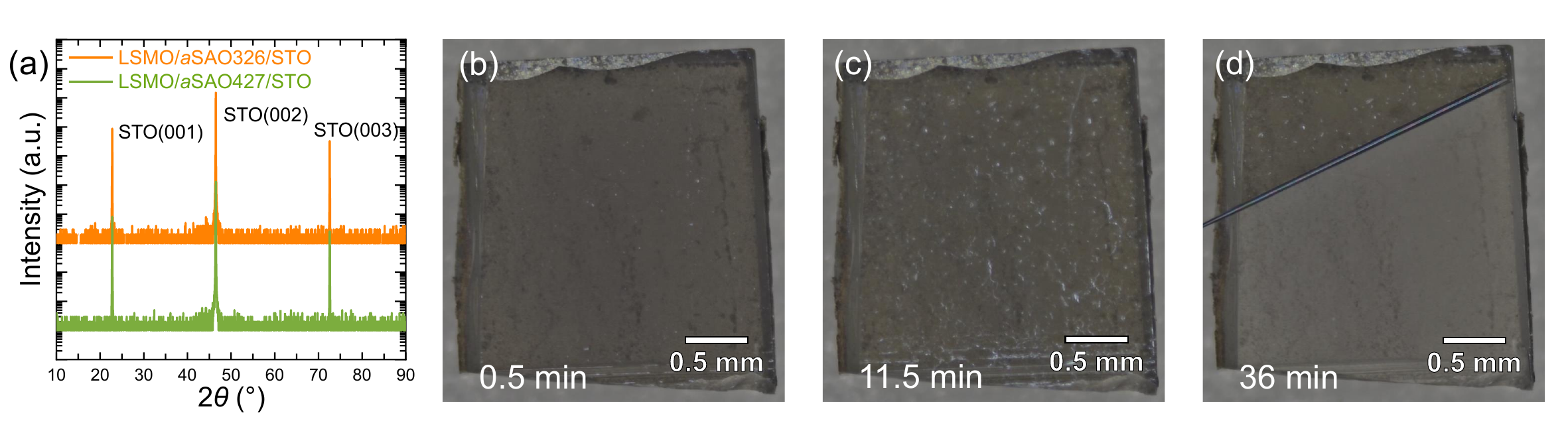}
	\caption{(a) XRD $2\theta$--$\omega$ scans of typical LSMO/$a$SAO326/SrTiO$_3$(001) and LSMO/$a$SAO437/SrTiO$_3$(001) heterostructures. The absence of peaks, aside from those associated with the SrTiO$_3$ substrate, suggests the amorphous nature of both the SAO and LSMO. (b-d) Optical microscopy images of an approximately $2.5 \times 2.5$ mm LSMO/$a$SAO326/SrTiO$_3$(001) heterostructure, taken at different time points following immersion in deionized water at ambient temperature (26.4 \textdegree C).}
	\label{SF2}
\end{figure}

\end{document}